\documentclass[aps,prl,twocolumn,superscriptaddress,showpacs,amsmath,amssymb]{revtex4-1}

\usepackage{graphicx}
\usepackage{dcolumn}
\usepackage{bm}
\usepackage{subfigure}
\usepackage{tabularx}
\usepackage{color}
\usepackage{hhline}

\begin{document}

\title{Gender differences in scientific collaborations: Women are more egalitarian than men}

\author{Eduardo B. Ara\'ujo}
\affiliation{Departamento de F\'isica, Universidade Federal do Cear\'a, Campus
do Pici 60451-970 Fortaleza, Cear\'a, Brazil}

\affiliation{Instituto Federal de Educa\c{c}\~{a}o, Ci\^{e}ncia e Tecnologia do
Cear\'a, Campus Acara\'u 62580-000 Acara\'u, Cear\'a, Brazil}

\author{Nuno A. M. Ara\'ujo}
\affiliation{Centro de F\'isica Te\'orica e Computacional, Faculdade de Ci\^{e}ncias, Universidade de Lisboa, Campo Grande,
1749-016 Lisboa, Portugal}

\affiliation{Departamento de F\'isica, Faculdade de Ci\^{e}ncias, Universidade de Lisboa, Campo Grande,
1749-016 Lisboa, Portugal}

\author{Andr\'e A. Moreira}
\affiliation{Departamento de F\'isica, Universidade Federal do Cear\'a, Campus
do Pici 60451-970 Fortaleza, Cear\'a, Brazil}

\author{Hans J. Herrmann}
\affiliation{Departamento de F\'isica, Universidade Federal do Cear\'a, Campus
do Pici 60451-970 Fortaleza, Cear\'a, Brazil}

\affiliation{Computational Physics for Engineering Materials, IfB, ETH Zurich,
Wolfgang-Pauli-Strasse 27, CH-8093 Zurich, Switzerland}

\author{Jos\'e S. Andrade Jr.}

\affiliation{Departamento de F\'isica, Universidade Federal do Cear\'a, Campus
do Pici 60451-970 Fortaleza, Cear\'a, Brazil}

\begin{abstract}
By analyzing a unique dataset of more than 270,000 scientists, we discovered
substantial gender differences in scientific collaborations.  While men are
more likely to collaborate with other men, women are more egalitarian. This is
consistently observed over all fields and regardless of the number of
collaborators a scientist has. The only exception is observed in the field of
engineering, where this gender bias disappears with increasing number of
collaborators. We also found that the distribution of the number of
collaborators follows a truncated power law with a cut-off that is gender
dependent and related to the gender differences in the number of published
papers. Considering interdisciplinary research, our analysis shows that men and
women behave similarly across fields, except in the case of natural sciences,
where women with many collaborators are more likely to have collaborators from
other fields.
\end{abstract}

\maketitle

\section{Introduction}
The challenges faced by women in academia are considered to be responsible for
their ubiquitous underrepresentation~\cite{leslie1996women, handelsman2005more,
schiebinger2007getting, duch2012possible}. Signs of gender asymmetries are
reported in several academic related activities such as
hiring~\cite{racusin2012science}, grant funding~\cite{boyle2015gender},
collaboration strategies~\cite{bozeman2011men}, and even in the ordering of the
list of authors in papers~\cite{west2013role}. These studies are usually based
on indirect analysis of scientific productivity~\cite{cole1984productivity} and
the evaluation of their career strategy~\cite{fox2001women, bozeman2011men,
duch2012possible}.  Here, we address the question of gender asymmetry from a
different perspective. Many successful and high-impact research works result
from the combination of skills, methods, and ideas of distinct team members.
Thus the mechanisms of team building strongly affect the collaboration network
structure and, consequently, its performance~\cite{Guimera2005}.  It is under
this framework that we analyze a dataset with more than 270,000 scientists in
Brazil and find that men are more likely to collaborate with other men than one
would expect from the gender distribution across fields. By contrast, for women
no significant gender bias is observed. 

In order to apply for grants and fellowships at any career level, scientists in
Brazil are required to register in the Lattes Platform~\cite{lattes}. This
results in a very detailed public database, which includes all active
scientists in Brazil and their full list of scientific publications. In
contrast with other databases, in this platform, articles are uniquely
identified by their DOI and possible ambiguities related to author names are
practically solved~\cite{newman2002scientific,araujo2014collaboration}.
Besides, it also includes personal information such as gender, research field,
and actual and previous academic positions. As a consequence, the application
of network science methods~\cite{rhoten2007women, szell2013how,
bottcher2016gender, cole1992making} to the collaboration network can provide
quantitative information for future discussions on the mechanisms responsible
for the observed gender disparities~\cite{prpic2002gender}.

The manuscript is organized as follows. In the next section, we present the
results for the distribution of the number of collaborators, gender
differences, and degree of research interdisciplinarity.  Final conclusions and
details about the methods are discussed afterwards.

\section{Results}
\begin{figure}
\begin{center}
\includegraphics[width=\columnwidth]{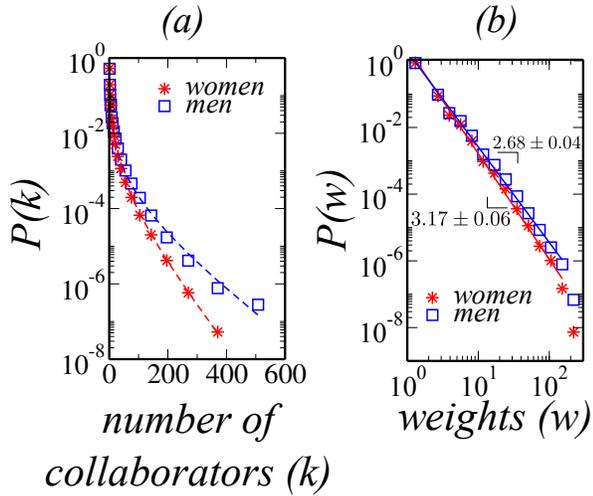}
\caption{ \label{fig:dist_weights} a) Distribution of the number of
collaborators for men (blue squares) and women (red asterisks). The
distributions are fitted with a truncated power law, $P(k) = A
k^{-\alpha}e^{-k/\beta}$, plotted as dashed lines with colors
corresponding to data points. The best fit is obtained for $\alpha =
1.53$, and $\beta = 85.4$ and $\beta = 49.5$, for men
and women, respectively.  b) Distribution of the number of recurrent
collaborations between scientists (weights) for men (blue squares) and
women (red asterisks). Solid lines are power-law fits, $P(w) =
Bw^{-\lambda}$, with colors corresponding to data points. For men,
$\lambda = 2.68\pm 0.04$, while for women $\lambda = 3.17\pm
0.06$. }
\end{center}
\end{figure}

The number of collaborators a scientist has is a cumulative quantity that
depends on the entire scientific career.  As shown in
Fig.~\ref{fig:dist_weights}a, the resulting distributions for men and women are
consistent with a truncated power law, $P= A k^{-\alpha} e^{-k/\beta}$, with
the same exponent $\alpha=1.53$ for both genders. However, the value of
the parameter $\beta$ for men is almost twice the one obtained for women,
namely, $85.4$ and $49.5$, respectively. This difference reflects
the tendency for men to have more collaborators than women.

In order to distinguish circumstantial from recurrent collaborations, we define
the weight $w$ of a collaboration between a pair of authors as the total number
of papers co-authored by them.  Figure~\ref{fig:dist_weights}b shows the
distribution of weights for both genders. The list-squares fit of the data to a
power law, $P(w)=B w^{-\lambda}$, gives $\lambda= 3.17\pm0.06$ and
$\lambda=2.68\pm0.04$, for men and women, respectively. This difference in
$\lambda$ might be related to the difference in the number of collaborators and
papers.  Table~\ref{tab:means} summarizes the average number of collaborators
and papers split by gender and research field. On average, men produce more
papers and have more collaborators than women even in the fields where women
are traditionally highly represented. An exception is the average number of
collaborators in Linguistics and Arts, which is very similar for both genders.

\begin{table}[b]
\caption{Mean number of collaborators and published papers for men
and women for each of the eight major fields: \textit{Agricultural Sciences}
(AGR), \textit{Applied Social Sciences} (SOC), \textit{Biological Sciences} (BIO), \textit{Exact and Earth
Sciences} (EXA), \textit{Humanities} (HUM), \textit{Health Sciences} (HEA), \textit{Engineering} (ENG) and
\textit{Linguistics and Arts} (LIN).
}
\begin{tabular}{|p{0.06\textwidth} | p{0.20\textwidth} | p{0.20\textwidth}|}
\hline
\textbf{Field} & \textbf{Mean number of collaborators (women/men)} & \textbf{Mean
number of papers (women/men)}\\
\hhline{|=|=|=|}
AGR & \hfill 9.20/13.6 &\hfill 9.45/17.5 \\
BIO & \hfill 10.9/14.9 &\hfill 10.0/17.7 \\
HEA & \hfill 7.65/11.2 &\hfill 9.18/17.7 \\
EXA & \hfill 7.90/9.84 &\hfill 9.49/15.6 \\
HUM & \hfill 3.16/3.31 &\hfill 7.54/11.4 \\
SOC & \hfill 2.85/3.57 &\hfill 6.62/10.5 \\
ENG & \hfill 6.02/6.50 &\hfill 8.08/11.0 \\
LIN & \hfill 2.06/2.04 &\hfill 8.25/11.2 \\
\hline
\end{tabular}
\label{tab:means}
\end{table}

\begin{figure}
\begin{center}
\includegraphics[width=\columnwidth]{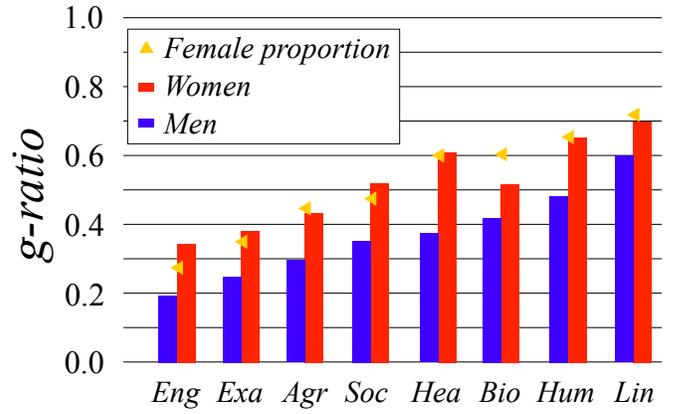}
\caption{ \label{fig:mean_g} Mean values of the {\it g-ratio} across fields. Same abreviations for the fields as in Table~\ref{tab:means}. Blue (left) and red (right)
bars represent values for men and women, respectively.  Yellow triangles show
the fraction of women working in the respective field.}
\end{center}
\end{figure}
\begin{figure}[b]
\begin{center}
\includegraphics[width=\columnwidth]{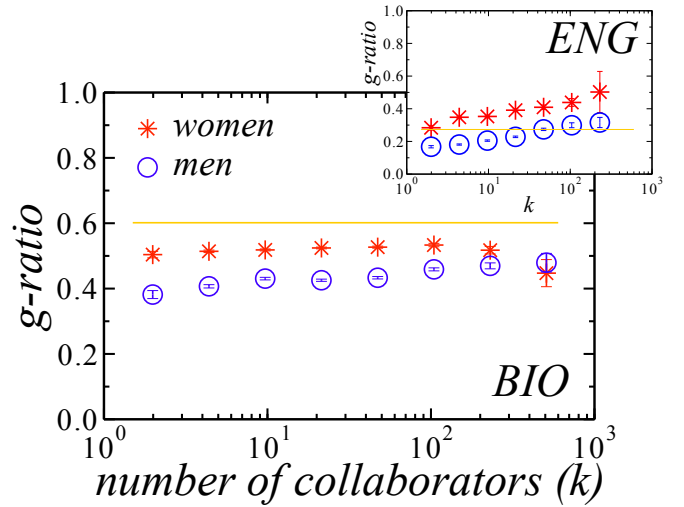}
\caption{ \label{fig:g_vs_k} Relation between the {\it g-ratio} and the number
of collaborators for \textit{Biological Sciences} (BIO, main plot) and \textit{Engineering} (ENG,
inset) for women (red stars) and men (blue circles).  Lines represent the
fraction of women in the respective field. Men are more likely to collaborate
with other men than with their female peers.  For \textit{Engineering}, the {\it
g-ratio} is even above the fraction of women in the field. Error bars indicate
the standard error for each bin.  }
\end{center}
\end{figure}
To evaluate homophily in the collaboration network, we define the gender ratio
of a scientist $i$, $\textit{g-ratio}_i$, as 
\begin{equation}
\textit{g-ratio}_{i}=\frac{\sum'_{j}w_{ij}}{\sum_{j} w_{ij}},
\end{equation}
where the sum in the denominator is over all collaborations, while the one in
the numerator is only over collaborations with women, and $w_{ij}$ is the
weight of the collaboration between scientist $i$ and $j$.
Figure~\ref{fig:mean_g} depicts the average {\it g-ratio} for men and women
across eight different fields. On average, women display a higher {\it
g-ratio}, regardless of the field, and always close to the fraction of women
working in the respective field. Men have relatively more collaborations with
other men, indicating a tendency to a homophilic pattern. Previous results
based on a rather small number of scientists suggested that women collaborate
more with other women~\cite{fox2001women, bozeman2004scientists}, but we show
that this is not the case for this much larger dataset.

The evidence of gender asymmetries raises the question of how it depends
on the number of collaborators $k$. Except for \textit{Engineering} (see inset in
Fig.~\ref{fig:g_vs_k}), the {\it g-ratio} does not depend strongly on
$k$. The values for women are always closer to the fraction of women
in the respective field and the values for men are always
consistently lower. This is exemplarily shown for \textit{Biological Sciences} in
Fig.~\ref{fig:g_vs_k}. For \textit{Engineering}, collaboration with women grows 
continuously with $k$ and even beyond the fraction of female scientists
for men and women with more collaborators (see
inset).
\begin{table} 
\caption{Average \textit{m-ratio} for men
and women for the eight major fields. Same abreviations for the fields as in Table~\ref{tab:means}. Error bars represent the standard error.}
\begin{tabular}{|l|r|r|}
\hline
{\bf Field} & {\bf Female} & {\bf Male} \\
\hhline{|=|=|=|}
AGR & 0.225 $\pm$ 0.002 & 0.198 $\pm$ 0.002 \\ 
BIO & 0.309 $\pm$ 0.002 & 0.292 $\pm$ 0.002 \\
HEA & 0.187 $\pm$ 0.002 & 0.167 $\pm$ 0.002 \\
EXA & 0.305 $\pm$ 0.003 & 0.252 $\pm$ 0.002 \\
HUM & 0.290 $\pm$ 0.003 & 0.332 $\pm$ 0.004 \\
SOC & 0.282 $\pm$ 0.004 & 0.240 $\pm$ 0.004 \\
ENG & 0.331 $\pm$ 0.005 & 0.254 $\pm$ 0.003 \\
LIN & 0.274 $\pm$ 0.007 & 0.303 $\pm$ 0.011 \\
\hline
\end{tabular}
\label{tab:m_ratio}
\end{table}
\begin{figure}[b]
\begin{center}
\includegraphics[width=\columnwidth]{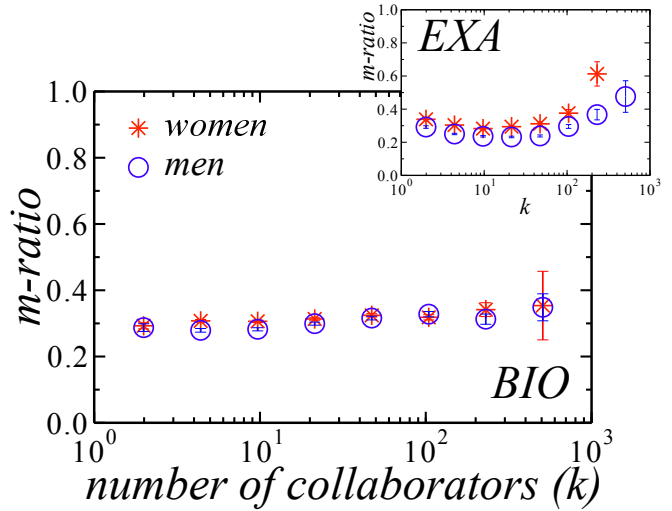}
\caption{ \label{fig:m_vs_k} Dependence of the {\it m-ratio} on the
number of collaborators in \textit{Biological Sciences} (BIO, main plot) and
\textit{Exact and Earth Sciences} (EXA, inset) for women (red starts) and men (blue
circles). Except for \textit{Exact and Earth Sciences} (see inset), there is only a slight difference
regarding multidisciplinary collaborations. Error bars indicate the
standard error for each bin. }
\end{center}
\end{figure}

It has been reported that women are more involved in interdisciplinary research
than their male peers~\cite{rhoten2007women}. To evaluate this tendency in our
dataset, we define the interdisciplinary ratio \textit{m-ratio}$_{i}$, as,
\begin{equation}
\textit{m-ratio}_{i}=\frac{\sum'_{j}w_{ij}}{\sum_{j} w_{ij}},
\end{equation}
where the sum in the denominator is over all collaborators, while the one in
the numerator is only over collaborators in a different field. The results are
summarized in Table~\ref{tab:m_ratio}. We observe that women have more
interdisciplinary collaborations than men for six fields, the exceptions being
the fields of \textit{Humanities} and \textit{Linguistics and Arts}. The
largest discrepancy is observed for \textit{Exact and Earth Sciences}.
Nonetheless, the differences are consistently smaller than the ones found for
the \textit{g-ratio}. When analyzing the dependence on the number of
collaborators, we observe the same tendency for men and women, as shown in
Fig.~\ref{fig:m_vs_k}. However, for \textit{Exact and Earth Sciences}, women
with a larger number of collaborators (more than 100) are considerably more
engaged in interdisciplinary research than men, with similar number of
collaborators (see inset).

\section{Conclusion}
We have found gender differences regarding scientific collaborations in
the Lattes Platform, a large dataset comprising more than 270,000
scientists. The number of collaborators and the weight of
collaborations, measured in terms of the number of common publications,
are both heavy tailed for men and women. Two metrics were introduced to
investigate gender differences, namely, the \textit{g-ratio}, that measures the
fraction of collaborations with women, and the \textit{m-ratio},
measuring the fraction of interdisciplinary collaborations. 

With the \textit{g-ratio}, we found that men collaborate more with other men
than with women, and this happens systematically across different fields and
regardless of their number of collaborators. The \textit{m-ratio} analysis
reveals that men and women have the same tendency to participate in
interdisciplinary research, with women being slightly more engaged. For \textit{Exact
and Earth Sciences}, women with a larger number of collaborators are
considerably more likely to work with scientists of a different field.

The path to gender balance in academia must involve not only government and
institutional support, but also consciousness of the asymmetries in the current
collaboration network. Our results are expected to provide quantitative support
to future analyses and discussions. The specific causes for the homophilic
pattern should also be investigated.

\section{Methods}
For analyzing the collaboration patterns in Lattes Platform, the XHTML
source code from circa 2.7 million curricula were extracted from the
website~\cite{lattes} in June 2012.  A parser was developed to extract
information from the downloaded information.

When filling their curricula, scientists may choose up to three
research fields. These are research topics organized in a hierarchical
tree-like structure comprising eight major fields: \textit{Agricultural Sciences}
(AGR), \textit{Applied Social Sciences} (SOC), \textit{Biological Sciences} (BIO), \textit{Exact}
and \textit{Earth Sciences} (EXA), \textit{Humanities} (HUM), \textit{Health Sciences} (HEA),
\textit{Engineering} (ENG) and \textit{Linguistics and Arts} (LIN). 
Each of these have their own subfields.  When assigning a
specific field to a scientist, we considered the first displayed
major field.

The procedure adopted to identify the collaborations is based on previous
studies of the Lattes Platform~\cite{menachalco2014brazilian}.  A list,
containing title, year of publication and number of authors of each paper
published is created. More than 3.5 million papers are present in this list.
The collaborations are identified by looking for duplicate records in the list.
Due to typographical errors \cite{o1993characteristics}, an exact string
matching would fail to identify collaborations. The Demearau-Levenshtein
approximate string matching algorithm~\cite{wagner1975extension} is therefore used to
define a distance between paper titles. Papers distant by less than 10\% of the
maximum distance, with the same number of authors and published in the same
year are considered to be the same. Due to the extensive number of records,
only papers published in the same year, with the same number of authors and
starting with the same letter are compared. Following this procedure, more than
620 thousand collaborations were identified.

With a list of duplicate papers, a bipartite network {\it BN} is
constructed containing two vertex classes, scientists ($R$) and papers
($P$). In this work, we analyze the projection of {\it BN} onto $R$,
which we call TCN (Total Collaboration Network). Two scientists are
said to collaborate if they are connected to a same paper in {\it BN}.
The weight $w_{ij}$ of their collaboration is defined as the number of
papers in {\it BN} which both are connected to. We note that these
networks are cumulative, with publications date spanning more
than five decades. 11.5\% of the scientists in TCN do not
include field information on their curricula. 
The proportions of female scientists varies across fields
\cite{west2013role} and the values for TCN are shown in Table
\ref{tab:proportion}.

\begin{table}
\caption{Number of scientists, fraction of the
Total Collaboration Network (TCN) they represent and proportion of women
for each of the eight major fields. The abbreviations are the same as in 
Table~\ref{tab:means}.}
\begin{tabular}{|p{0.06\textwidth} | p{0.25\textwidth} | r|}
\hline
{\bf Field} & {\bf Number of scientists (Fraction on TCN)} & {\bf Female Proportion} \\
\hhline{|=|=|=|}
AGR & \hfill 31812 (11.6\%)& 44.4\%\\
BIO & \hfill 39767 (14.5\%) & 60.1\%\\
HEA & \hfill 67561 (24.6\%) & 59.8\% \\
EXA & \hfill 33310 (12.1\%) & 34.7\%\\
HUM & \hfill 26263 (9.55\%) & 65.1\%\\
SOC & \hfill 20806 (7.57\%) & 47.3\%\\
ENG & \hfill 18365 (6.68\%) & 27.2\%\\
LIN & \hfill 5202 (1.90\%) & 71.6\%\\
\hline
\end{tabular}
\label{tab:proportion}
\end{table}

Previous works on gender and collaboration~\cite{kyvik1996child, fox2001women,
bozeman2004scientists, rhoten2007women, bozeman2011men} had information from a
much smaller number of authors, usually much less than 10,000. Here, we have
information concerning the productivity (as measured by article output) of
275,061 scientists with published papers on periodicals, 130,525 men (47.4\%)
and 144'440 women (52.5\%). Only 96 scientists do not display the gender
information on the curriculum.  90.4\% belong to the giant component of the TCN.

\textbf{Acknowledgments} 
We acknowledge financial support from the ETH Risk Center, the Brazilian
institute INCT-SC, and grant number 319968-FlowCCS of the European Researcher
Council. NA acknowledges financial support from the Portuguese Foundation for
Sciences and Technology (FCT) under Contracts nos. UID/FIS/00618/2013 and
IF/00255/2013.

\bibliography{gender2}

\begin{thebibliography}{24}%
\makeatletter
\providecommand \@ifxundefined [1]{%
 \@ifx{#1\undefined}
}%
\providecommand \@ifnum [1]{%
 \ifnum #1\expandafter \@firstoftwo
 \else \expandafter \@secondoftwo
 \fi
}%
\providecommand \@ifx [1]{%
 \ifx #1\expandafter \@firstoftwo
 \else \expandafter \@secondoftwo
 \fi
}%
\providecommand \natexlab [1]{#1}%
\providecommand \enquote  [1]{``#1''}%
\providecommand \bibnamefont  [1]{#1}%
\providecommand \bibfnamefont [1]{#1}%
\providecommand \citenamefont [1]{#1}%
\providecommand \href@noop [0]{\@secondoftwo}%
\providecommand \href [0]{\begingroup \@sanitize@url \@href}%
\providecommand \@href[1]{\@@startlink{#1}\@@href}%
\providecommand \@@href[1]{\endgroup#1\@@endlink}%
\providecommand \@sanitize@url [0]{\catcode `\\12\catcode `\$12\catcode
  `\&12\catcode `\#12\catcode `\^12\catcode `\_12\catcode `\%12\relax}%
\providecommand \@@startlink[1]{}%
\providecommand \@@endlink[0]{}%
\providecommand \url  [0]{\begingroup\@sanitize@url \@url }%
\providecommand \@url [1]{\endgroup\@href {#1}{\urlprefix }}%
\providecommand \urlprefix  [0]{URL }%
\providecommand \Eprint [0]{\href }%
\providecommand \doibase [0]{http://dx.doi.org/}%
\providecommand \selectlanguage [0]{\@gobble}%
\providecommand \bibinfo  [0]{\@secondoftwo}%
\providecommand \bibfield  [0]{\@secondoftwo}%
\providecommand \translation [1]{[#1]}%
\providecommand \BibitemOpen [0]{}%
\providecommand \bibitemStop [0]{}%
\providecommand \bibitemNoStop [0]{.\EOS\space}%
\providecommand \EOS [0]{\spacefactor3000\relax}%
\providecommand \BibitemShut  [1]{\csname bibitem#1\endcsname}%
\let\auto@bib@innerbib\@empty
\bibitem [{\citenamefont {Leslie}\ \emph {et~al.}(1996)\citenamefont {Leslie},
  \citenamefont {McClure},\ and\ \citenamefont {Oaxaca}}]{leslie1996women}%
  \BibitemOpen
  \bibfield  {author} {\bibinfo {author} {\bibfnamefont {L.~L.}\ \bibnamefont
  {Leslie}}, \bibinfo {author} {\bibfnamefont {G.~T.}\ \bibnamefont {McClure}},
  \ and\ \bibinfo {author} {\bibfnamefont {R.~L.}\ \bibnamefont {Oaxaca}},\
  }\href@noop {} {\bibfield  {journal} {\bibinfo  {journal} {The Journal of
  Higher Education}\ }\textbf {\bibinfo {volume} {69}},\ \bibinfo {pages} {239}
  (\bibinfo {year} {1996})}\BibitemShut {NoStop}%
\bibitem [{\citenamefont {Handelsman}\ \emph {et~al.}(2005)\citenamefont
  {Handelsman}, \citenamefont {Cantor}, \citenamefont {Carnes}, \citenamefont
  {Denton}, \citenamefont {Fine}, \citenamefont {Grosz}, \citenamefont
  {Hinshaw}, \citenamefont {Marrett}, \citenamefont {Rosser}, \citenamefont
  {Shalala} \emph {et~al.}}]{handelsman2005more}%
  \BibitemOpen
  \bibfield  {author} {\bibinfo {author} {\bibfnamefont {J.}~\bibnamefont
  {Handelsman}}, \bibinfo {author} {\bibfnamefont {N.}~\bibnamefont {Cantor}},
  \bibinfo {author} {\bibfnamefont {M.}~\bibnamefont {Carnes}}, \bibinfo
  {author} {\bibfnamefont {D.}~\bibnamefont {Denton}}, \bibinfo {author}
  {\bibfnamefont {E.}~\bibnamefont {Fine}}, \bibinfo {author} {\bibfnamefont
  {B.}~\bibnamefont {Grosz}}, \bibinfo {author} {\bibfnamefont
  {V.}~\bibnamefont {Hinshaw}}, \bibinfo {author} {\bibfnamefont
  {C.}~\bibnamefont {Marrett}}, \bibinfo {author} {\bibfnamefont
  {S.}~\bibnamefont {Rosser}}, \bibinfo {author} {\bibfnamefont
  {D.}~\bibnamefont {Shalala}},  \emph {et~al.},\ }\href@noop {} {\bibfield
  {journal} {\bibinfo  {journal} {Science}\ }\textbf {\bibinfo {volume}
  {309}},\ \bibinfo {pages} {1190} (\bibinfo {year} {2005})}\BibitemShut
  {NoStop}%
\bibitem [{\citenamefont {Schiebinger}(2007)}]{schiebinger2007getting}%
  \BibitemOpen
  \bibfield  {author} {\bibinfo {author} {\bibfnamefont {L.}~\bibnamefont
  {Schiebinger}},\ }\href@noop {} {\bibfield  {journal} {\bibinfo  {journal}
  {Harvard Journal of Law \& Gender}\ }\textbf {\bibinfo {volume} {30}},\
  \bibinfo {pages} {350} (\bibinfo {year} {2007})}\BibitemShut {NoStop}%
\bibitem [{\citenamefont {Duch}\ \emph {et~al.}(2012)\citenamefont {Duch},
  \citenamefont {Zeng}, \citenamefont {Sales-Pardo}, \citenamefont {Radicchi},
  \citenamefont {Otis}, \citenamefont {Woodruff},\ and\ \citenamefont
  {Amaral}}]{duch2012possible}%
  \BibitemOpen
  \bibfield  {author} {\bibinfo {author} {\bibfnamefont {J.}~\bibnamefont
  {Duch}}, \bibinfo {author} {\bibfnamefont {X.~H.~T.}\ \bibnamefont {Zeng}},
  \bibinfo {author} {\bibfnamefont {M.}~\bibnamefont {Sales-Pardo}}, \bibinfo
  {author} {\bibfnamefont {F.}~\bibnamefont {Radicchi}}, \bibinfo {author}
  {\bibfnamefont {S.}~\bibnamefont {Otis}}, \bibinfo {author} {\bibfnamefont
  {T.~K.}\ \bibnamefont {Woodruff}}, \ and\ \bibinfo {author} {\bibfnamefont
  {L.~A.~N.}\ \bibnamefont {Amaral}},\ }\href@noop {} {\bibfield  {journal}
  {\bibinfo  {journal} {PLOS One}\ }\textbf {\bibinfo {volume} {7}},\ \bibinfo
  {pages} {e51332} (\bibinfo {year} {2012})}\BibitemShut {NoStop}%
\bibitem [{\citenamefont {Moss-Racusin}\ \emph {et~al.}(2012)\citenamefont
  {Moss-Racusin}, \citenamefont {Dovidio}, \citenamefont {Crescoll},
  \citenamefont {Graham},\ and\ \citenamefont
  {Handelsman}}]{racusin2012science}%
  \BibitemOpen
  \bibfield  {author} {\bibinfo {author} {\bibfnamefont {C.~A.}\ \bibnamefont
  {Moss-Racusin}}, \bibinfo {author} {\bibfnamefont {J.~F.}\ \bibnamefont
  {Dovidio}}, \bibinfo {author} {\bibfnamefont {V.~L.}\ \bibnamefont
  {Crescoll}}, \bibinfo {author} {\bibfnamefont {M.~J.}\ \bibnamefont
  {Graham}}, \ and\ \bibinfo {author} {\bibfnamefont {J.}~\bibnamefont
  {Handelsman}},\ }\href@noop {} {\bibfield  {journal} {\bibinfo  {journal}
  {Proceedings of the National Academy of Sciences}\ }\textbf {\bibinfo
  {volume} {109}},\ \bibinfo {pages} {16474} (\bibinfo {year}
  {2012})}\BibitemShut {NoStop}%
\bibitem [{\citenamefont {Boyle}\ \emph {et~al.}(2015)\citenamefont {Boyle},
  \citenamefont {Smith}, \citenamefont {N.~J.~Cooper},\ and\ \citenamefont
  {O'Connor}}]{boyle2015gender}%
  \BibitemOpen
  \bibfield  {author} {\bibinfo {author} {\bibfnamefont {P.~J.}\ \bibnamefont
  {Boyle}}, \bibinfo {author} {\bibfnamefont {L.~K.}\ \bibnamefont {Smith}},
  \bibinfo {author} {\bibfnamefont {K.~S.~W.}\ \bibnamefont {N.~J.~Cooper}}, \
  and\ \bibinfo {author} {\bibfnamefont {H.}~\bibnamefont {O'Connor}},\
  }\href@noop {} {\bibfield  {journal} {\bibinfo  {journal} {Nature}\ }\textbf
  {\bibinfo {volume} {525}},\ \bibinfo {pages} {181} (\bibinfo {year}
  {2015})}\BibitemShut {NoStop}%
\bibitem [{\citenamefont {Bozeman}\ and\ \citenamefont
  {Gaughan}(2011)}]{bozeman2011men}%
  \BibitemOpen
  \bibfield  {author} {\bibinfo {author} {\bibfnamefont {B.}~\bibnamefont
  {Bozeman}}\ and\ \bibinfo {author} {\bibfnamefont {M.}~\bibnamefont
  {Gaughan}},\ }\href@noop {} {\bibfield  {journal} {\bibinfo  {journal}
  {Research Policy}\ }\textbf {\bibinfo {volume} {40}},\ \bibinfo {pages}
  {1393} (\bibinfo {year} {2011})}\BibitemShut {NoStop}%
\bibitem [{\citenamefont {West}\ \emph {et~al.}(2013)\citenamefont {West},
  \citenamefont {Jacquet}, \citenamefont {King}, \citenamefont {Correll},\ and\
  \citenamefont {Bergstrom}}]{west2013role}%
  \BibitemOpen
  \bibfield  {author} {\bibinfo {author} {\bibfnamefont {J.}~\bibnamefont
  {West}}, \bibinfo {author} {\bibfnamefont {J.}~\bibnamefont {Jacquet}},
  \bibinfo {author} {\bibfnamefont {M.~M.}\ \bibnamefont {King}}, \bibinfo
  {author} {\bibfnamefont {S.~J.}\ \bibnamefont {Correll}}, \ and\ \bibinfo
  {author} {\bibfnamefont {C.~T.}\ \bibnamefont {Bergstrom}},\ }\href@noop {}
  {\bibfield  {journal} {\bibinfo  {journal} {PLOS One}\ }\textbf {\bibinfo
  {volume} {8}},\ \bibinfo {pages} {e66121} (\bibinfo {year}
  {2013})}\BibitemShut {NoStop}%
\bibitem [{\citenamefont {Cole}\ and\ \citenamefont
  {Zuckerman}(1984)}]{cole1984productivity}%
  \BibitemOpen
  \bibfield  {author} {\bibinfo {author} {\bibfnamefont {J.~R.}\ \bibnamefont
  {Cole}}\ and\ \bibinfo {author} {\bibfnamefont {H.}~\bibnamefont
  {Zuckerman}},\ }\href@noop {} {\emph {\bibinfo {title} {The productivity
  puzzle: persistence and change in patterns of publication of men and women
  scientists}}}\ (\bibinfo  {publisher} {JAI Press},\ \bibinfo {address}
  {Greenwich},\ \bibinfo {year} {1984})\ pp.\ \bibinfo {pages}
  {217--258}\BibitemShut {NoStop}%
\bibitem [{\citenamefont {Fox}(2001)}]{fox2001women}%
  \BibitemOpen
  \bibfield  {author} {\bibinfo {author} {\bibfnamefont {M.}~\bibnamefont
  {Fox}},\ }\href@noop {} {\bibfield  {journal} {\bibinfo  {journal} {Gender \&
  Society}\ }\textbf {\bibinfo {volume} {15}},\ \bibinfo {pages} {654}
  (\bibinfo {year} {2001})}\BibitemShut {NoStop}%
\bibitem [{\citenamefont {Guimer\`a}\ \emph {et~al.}(2005)\citenamefont
  {Guimer\`a}, \citenamefont {Uzzi}, \citenamefont {Spiro},\ and\ \citenamefont
  {Amaral}}]{Guimera2005}%
  \BibitemOpen
  \bibfield  {author} {\bibinfo {author} {\bibfnamefont {R.}~\bibnamefont
  {Guimer\`a}}, \bibinfo {author} {\bibfnamefont {B.}~\bibnamefont {Uzzi}},
  \bibinfo {author} {\bibfnamefont {J.}~\bibnamefont {Spiro}}, \ and\ \bibinfo
  {author} {\bibfnamefont {L.~A.~N.}\ \bibnamefont {Amaral}},\ }\href@noop {}
  {\bibfield  {journal} {\bibinfo  {journal} {Science}\ }\textbf {\bibinfo
  {volume} {308}},\ \bibinfo {pages} {5722} (\bibinfo {year}
  {2005})}\BibitemShut {NoStop}%
\bibitem [{lat()}]{lattes}%
  \BibitemOpen
  \href@noop {} {\enquote {\bibinfo {title} {Lattes platform},}\ }\bibinfo
  {howpublished} {\url{http://lattes.cnpq.br}}\BibitemShut {NoStop}%
\bibitem [{\citenamefont {Newman}(2002)}]{newman2002scientific}%
  \BibitemOpen
  \bibfield  {author} {\bibinfo {author} {\bibfnamefont {M.~E.~J.}\
  \bibnamefont {Newman}},\ }\href@noop {} {\bibfield  {journal} {\bibinfo
  {journal} {Phys Rev E Stat Nonlin Soft Matter Phys}\ }\textbf {\bibinfo
  {volume} {64}},\ \bibinfo {pages} {016131} (\bibinfo {year}
  {2002})}\BibitemShut {NoStop}%
\bibitem [{\citenamefont {Ara{\'u}jo}\ \emph {et~al.}(2014)\citenamefont
  {Ara{\'u}jo}, \citenamefont {Moreira}, \citenamefont {Furtado}, \citenamefont
  {Pequeno},\ and\ \citenamefont {Andrade}}]{araujo2014collaboration}%
  \BibitemOpen
  \bibfield  {author} {\bibinfo {author} {\bibfnamefont {E.~B.}\ \bibnamefont
  {Ara{\'u}jo}}, \bibinfo {author} {\bibfnamefont {A.~A.}\ \bibnamefont
  {Moreira}}, \bibinfo {author} {\bibfnamefont {V.}~\bibnamefont {Furtado}},
  \bibinfo {author} {\bibfnamefont {T.~H.~C.}\ \bibnamefont {Pequeno}}, \ and\
  \bibinfo {author} {\bibfnamefont {J.~S.}\ \bibnamefont {Andrade},
  \bibfnamefont {Jr}},\ }\href@noop {} {\bibfield  {journal} {\bibinfo
  {journal} {Plos One}\ }\textbf {\bibinfo {volume} {9}},\ \bibinfo {pages}
  {e90537} (\bibinfo {year} {2014})}\BibitemShut {NoStop}%
\bibitem [{\citenamefont {Rhoten}\ and\ \citenamefont
  {Pfirman}(2007)}]{rhoten2007women}%
  \BibitemOpen
  \bibfield  {author} {\bibinfo {author} {\bibfnamefont {D.}~\bibnamefont
  {Rhoten}}\ and\ \bibinfo {author} {\bibfnamefont {S.}~\bibnamefont
  {Pfirman}},\ }\href@noop {} {\bibfield  {journal} {\bibinfo  {journal}
  {Research policy}\ }\textbf {\bibinfo {volume} {36}},\ \bibinfo {pages} {56}
  (\bibinfo {year} {2007})}\BibitemShut {NoStop}%
\bibitem [{\citenamefont {Szell}\ and\ \citenamefont
  {Thurner}(2013)}]{szell2013how}%
  \BibitemOpen
  \bibfield  {author} {\bibinfo {author} {\bibfnamefont {M.}~\bibnamefont
  {Szell}}\ and\ \bibinfo {author} {\bibfnamefont {S.}~\bibnamefont
  {Thurner}},\ }\href@noop {} {\bibfield  {journal} {\bibinfo  {journal}
  {Scientific Reports}\ }\textbf {\bibinfo {volume} {3}},\ \bibinfo {pages} {1}
  (\bibinfo {year} {2013})}\BibitemShut {NoStop}%
\bibitem [{\citenamefont {B{\"o}ttcher}\ \emph {et~al.}(2016)\citenamefont
  {B{\"o}ttcher}, \citenamefont {Ara{\'u}jo}, \citenamefont {Nagler},
  \citenamefont {Mendes}, \citenamefont {Helbing},\ and\ \citenamefont
  {Herrmann}}]{bottcher2016gender}%
  \BibitemOpen
  \bibfield  {author} {\bibinfo {author} {\bibfnamefont {L.}~\bibnamefont
  {B{\"o}ttcher}}, \bibinfo {author} {\bibfnamefont {N.~A.~M.}\ \bibnamefont
  {Ara{\'u}jo}}, \bibinfo {author} {\bibfnamefont {J.}~\bibnamefont {Nagler}},
  \bibinfo {author} {\bibfnamefont {J.~F.~F.}\ \bibnamefont {Mendes}}, \bibinfo
  {author} {\bibfnamefont {D.}~\bibnamefont {Helbing}}, \ and\ \bibinfo
  {author} {\bibfnamefont {H.~J.}\ \bibnamefont {Herrmann}},\ }\href@noop {}
  {\bibfield  {journal} {\bibinfo  {journal} {PLoS One}\ }\textbf {\bibinfo
  {volume} {11}},\ \bibinfo {pages} {e0149514} (\bibinfo {year}
  {2016})}\BibitemShut {NoStop}%
\bibitem [{\citenamefont {Cole}(1992)}]{cole1992making}%
  \BibitemOpen
  \bibfield  {author} {\bibinfo {author} {\bibfnamefont {S.}~\bibnamefont
  {Cole}},\ }\href@noop {} {\emph {\bibinfo {title} {Making Science: Between
  Nature and Scociety}}}\ (\bibinfo  {publisher} {Harvard University Press},\
  \bibinfo {address} {Cambridge},\ \bibinfo {year} {1992})\BibitemShut
  {NoStop}%
\bibitem [{\citenamefont {Prpi{\'c }}(2002)}]{prpic2002gender}%
  \BibitemOpen
  \bibfield  {author} {\bibinfo {author} {\bibfnamefont {K.}~\bibnamefont
  {Prpi{\'c }}},\ }\href@noop {} {\bibfield  {journal} {\bibinfo  {journal}
  {Scientometrics}\ }\textbf {\bibinfo {volume} {55}},\ \bibinfo {pages} {27}
  (\bibinfo {year} {2002})}\BibitemShut {NoStop}%
\bibitem [{\citenamefont {Bozeman}\ and\ \citenamefont
  {Corley}(2004)}]{bozeman2004scientists}%
  \BibitemOpen
  \bibfield  {author} {\bibinfo {author} {\bibfnamefont {B.}~\bibnamefont
  {Bozeman}}\ and\ \bibinfo {author} {\bibfnamefont {E.}~\bibnamefont
  {Corley}},\ }\href@noop {} {\bibfield  {journal} {\bibinfo  {journal}
  {Research Policy}\ }\textbf {\bibinfo {volume} {33}},\ \bibinfo {pages} {599}
  (\bibinfo {year} {2004})}\BibitemShut {NoStop}%
\bibitem [{\citenamefont {Mena-Chalco}\ \emph {et~al.}(2014)\citenamefont
  {Mena-Chalco}, \citenamefont {Digiampietri}, \citenamefont {Lopes},\ and\
  \citenamefont {Jr.}}]{menachalco2014brazilian}%
  \BibitemOpen
  \bibfield  {author} {\bibinfo {author} {\bibfnamefont {J.~P.}\ \bibnamefont
  {Mena-Chalco}}, \bibinfo {author} {\bibfnamefont {L.~A.}\ \bibnamefont
  {Digiampietri}}, \bibinfo {author} {\bibfnamefont {F.~M.}\ \bibnamefont
  {Lopes}}, \ and\ \bibinfo {author} {\bibfnamefont {R.~M.~C.}\ \bibnamefont
  {Jr.}},\ }\href@noop {} {\bibfield  {journal} {\bibinfo  {journal} {Journal
  of the Association for Information Science and Technology}\ }\textbf
  {\bibinfo {volume} {65}},\ \bibinfo {pages} {1424} (\bibinfo {year}
  {2014})}\BibitemShut {NoStop}%
\bibitem [{\citenamefont {O'Neill}\ \emph {et~al.}(1993)\citenamefont
  {O'Neill}, \citenamefont {Rogers},\ and\ \citenamefont
  {Oskins}}]{o1993characteristics}%
  \BibitemOpen
  \bibfield  {author} {\bibinfo {author} {\bibfnamefont {E.~T.}\ \bibnamefont
  {O'Neill}}, \bibinfo {author} {\bibfnamefont {S.~A.}\ \bibnamefont {Rogers}},
  \ and\ \bibinfo {author} {\bibfnamefont {W.~M.}\ \bibnamefont {Oskins}},\
  }\href@noop {} {\bibfield  {journal} {\bibinfo  {journal} {Libr. Resour.
  Tech. Serv.}\ }\textbf {\bibinfo {volume} {37}},\ \bibinfo {pages} {59}
  (\bibinfo {year} {1993})}\BibitemShut {NoStop}%
\bibitem [{\citenamefont {Wagner}\ and\ \citenamefont
  {Lowrance}(1975)}]{wagner1975extension}%
  \BibitemOpen
  \bibfield  {author} {\bibinfo {author} {\bibfnamefont {R.~A.}\ \bibnamefont
  {Wagner}}\ and\ \bibinfo {author} {\bibfnamefont {R.}~\bibnamefont
  {Lowrance}},\ }\href@noop {} {\bibfield  {journal} {\bibinfo  {journal} {J.
  Assoc. Comput. Mach.}\ }\textbf {\bibinfo {volume} {22}},\ \bibinfo {pages}
  {177} (\bibinfo {year} {1975})}\BibitemShut {NoStop}%
\bibitem [{\citenamefont {Kyvik}\ and\ \citenamefont
  {Teigen}(1996)}]{kyvik1996child}%
  \BibitemOpen
  \bibfield  {author} {\bibinfo {author} {\bibfnamefont {S.}~\bibnamefont
  {Kyvik}}\ and\ \bibinfo {author} {\bibfnamefont {M.}~\bibnamefont {Teigen}},\
  }\href@noop {} {\bibfield  {journal} {\bibinfo  {journal} {Science,
  Technology \& Human Values}\ }\textbf {\bibinfo {volume} {21}},\ \bibinfo
  {pages} {54} (\bibinfo {year} {1996})}\BibitemShut {NoStop}%
\end{thebibliography}%

\end{document}